\documentclass[a4paper,12pt]{elsarticle}

\usepackage{amssymb}
\usepackage{amsmath}
\usepackage[utf8]{inputenc}
\usepackage[ngerman, french, english]{babel}
\usepackage{caption}
\usepackage{subcaption}
\usepackage{gensymb}
\usepackage{lineno}
\usepackage{xcolor}
\usepackage{hyperref}
\hypersetup{
    colorlinks=false,
    hidelinks=true,
}

\journal{Physics of the Dark Universe}

\begin{document}

\begin{frontmatter}



\title{Inverse Compton emission from heavy WIMP annihilations in the Galactic Centre}

\author[MPIK,IFT]{Julia I. Djuvsland\corref{cor1}}
\ead{julia.djuvsland@uib.no}
\author[MPIK]{Jim Hinton}
\author[MPIK]{Brian Reville}
\cortext[cor1]{Corresponding author}
\affiliation[MPIK]{organization={Max-Planck-Institut f\"ur Kernphysik},
            addressline={Saupfercheckweg 1}, 
            city={Heidelberg},
            postcode={69117}, 
            country={Germany}}
\affiliation[IFT]{organization={Department for Physics and Technology, University of Bergen},
            addressline={Allegaten 55}, 
            city={Bergen},
            postcode={5020}, 
            country={Norway}}
\begin{abstract}
A thermal relic WIMP remains a prime candidate for the nature of Dark Matter, particularly for the more poorly constrained case of a heavy ($\gtrsim$ 1 TeV) WIMP. The highest fluxes from WIMP annihilations are expected in the region of the Galactic Centre (GC) where current and near future gamma-ray observatories can be exploited to place tight limits on the WIMP paradigm. 
It is regularly noted that the annihilation flux of gammas will be accompanied by charged secondary particles which can produce `delayed' inverse Compton (IC) gamma-ray emission, but this component is often neglected in indirect Dark Matter searches. In this work the inverse Compton emission is studied for the specific conditions of heavy WIMP annihilation in the GC. Using models for the magnetic and radiation fields of the region, and taking into consideration the transport of secondary particles, we find that for TeV WIMPs the IC component cannot be neglected in the GC, with the particles produced cooling within the region rather than propagating out in to the Galaxy. This effect changes the predicted spectral shape substantially and thus boosts the detection prospects for heavy WIMPs.
\end{abstract}




\end{frontmatter}

%\linenumbers


\section{Introduction}
\label{sec:intro}
A weakly interacting massive particle (WIMP), left over as a \emph{thermal relic}~\cite{Arcadi_2018} of the Big Bang, remains a compelling candidate for the Dark Matter (DM) that makes up 27\% of the energy density in our universe~\cite{dmfrac}.
While there is ample evidence for the existence of DM through astrophysical observations like gravitational lensing~\cite{Massey_2007}, galaxy rotation curves~\cite{Sofue_2001} and the study of the Cosmic Microwave Background (CMB)~\cite{dmfrac}, the particle nature of DM is still poorly understood. To address this, many efforts are currently underway to search for a DM particle candidate at particle colliders, in underground laboratories and by exploring the gamma-ray sky -- to name just a few. 

A popular model explored in this context is that of annihilating WIMPs, whose parameter space is narrowed by ongoing searches. The DM searches at CERN's Large Hadron Collider (LHC) explore the GeV WIMP mass region and seem to advocate heavy candidates, e.g.~\cite{Aad_2022, CMS-EXO-19-003}. This mass range can also be explored via indirect DM searches that measure gamma-rays arriving at Earth using detectors such as the Fermi-LAT satellite, ground-based Imaging Atmospheric Cherenkov Telescopes (IACTs) or high altitude air shower observatories. 

While the current and future IACTs - H.E.S.S.~\cite{hess_22}, MAGIC~\cite{magic_22}, VERITAS~\cite{Archambault_2017}  and CTA~\cite{Acharyya_2021} -- are most sensitive to WIMPs with TeV masses, high altitude air shower observatories such as HAWC, LHAASO and SWGO can constrain even higher WIMP masses ($O$(100\,TeV))~\cite{HAWC, LHAASO, Viana_2019}.
Fermi-LAT currently quotes its most stringent exclusion limits in the GeV range with little sensitivity for TeV-WIMPs~\cite{Fermi_2017}. These limits are however obtained by only taking the \emph{direct} photon component into account. With direct photons, we refer to photons that are directly produced during the WIMP annihilation and the consecutive decay of the immediate annihilation products, i.e. during final-state radiation and hadronisation processes. Yet, gamma-rays can also be produced indirectly, in particular by the \emph{direct} electrons\footnote{For simplicity the term electron is used to describe both direct electrons and direct positrons arising from WIMP annihilation throughout this paper.} that arise from the WIMP annihilation. They give rise to energetic photons via inverse Compton (IC) scattering of ambient photons, and also synchrotron and Bremsstrahlung emission. These \emph{indirect} photons, from the IC process in particular, contribute to the expected gamma-ray signal from WIMPs and have typically lower energies than the signal component from the direct photons. 

The shape and intensity of the indirect photon component of the Galactic WIMP gamma-ray spectrum depends highly on the transport properties of charged particles near the GC. This is because the GC conditions impact the lifetime of the radiating electrons. 
Cosmic-ray (CR) transport around the Galaxy remains however an unsolved problem. The majority of observational constraints are inferred from local measurements, which alone can not provide an accurate picture of the relevant physical parameters in the GC region. Gamma-ray observations provide an indirect probe of the CR distribution, and provide compelling evidence for greatly reduced diffusion coefficients with respect to the Galactic average \cite{HESSPevatron} in this region.
The complex magnetic field topology \cite{Ferriere09,Guendez} and fluid motions (in particular outflows \cite{BlandHawthorn03,Ponti19}) in the GC region influence the emission profiles of both the astrophysical foregrounds and crucially, any delayed emission probed in indirect WIMP searches. The transport of the emitting particles in this extreme environment are not expected to match those of the Galactic average: using local CR observations one would infer a scattering mean free path of approximately 10 pc for particles with TeV energies. In this work we highlight the physical motivation for reduced scattering lengths in the inner kpc region. The results that follow make the further simplifying assumption of isotropic diffusion.

The importance of taking the Bremsstrahlung and IC-photons into account in indirect WIMP searches was discussed previously, e.g.~\cite{Cholis_2009, COOKBOOK}, especially for light WIMP masses~\cite{Cirelli_2013} and in the context of the Fermi GeV excess~\cite{Lacroix_2014}. Also the importance of synchrotron radiation effects were studied~\cite{Zhang_2009, COOKBOOK}. In~\cite{Egorov_2016} the potential of a multiwavelength analysis combining the gamma ray and microwave observations was explored.

Still, the indirect component of the gamma-ray signal is frequently ignored in indirect DM searches. Therefore this work highlights the fact that the indirect component is not negligible for TeV mass WIMP annihilation in the Galactic Centre. As the indirect component leads to additional photons (with lower energies) than expected from the direct photon contribution only, ignoring this contribution would result in searches underestimating their sensitivity and/or misinterpretation of a detected signal.

\section{Cosmic ray transport and energy losses in the GC region}
\label{sec:transport}
The central few hundred pc of our Galaxy is a complex and unique environment, dominated observationally by the Central Molecular Zone (CMZ), with \(\sim5 \times10^{7}\) solar masses of gas in a thin ($\sim$30 pc) highly structured disc. Densities in the CMZ are orders of magnitude larger than is typical in the disc of the Milky Way (MW) and strong magnetic fields have been detected~\cite{Ferriere09}. Either the star formation of the CMZ, or the central supermassive black hole Sgr A$^\star$ is responsible for blowing the Fermi/eROSITA bubbles~\cite{FermiBubbles,eROSITABubbles}, which extend for 55$^\circ$/80$^\circ$ degrees above and below the GC, and which appear to be connected to the CMZ by \emph{Chimneys} recently detected in the X-ray band~\cite{Ponti19}. The heart of the Galaxy is also naturally where the most intense large scale radiation fields can be found \cite{radiationField}. Taken together it is clear that the GC is a unique environment, in no way resembling typical MW conditions.  

Whilst emission from the entire GC region is expected from the annihilation of a hypothetical thermal relic WIMP, the emission within the CMZ itself would be very difficult to disentangle from the foreground astrophysical emission~\cite{HESSPevatron, HESS2018,MagicGC,VeritasGC}. Above and below the CMZ, astrophysical foregrounds are greatly reduced, and conditions are expected to be much less extreme than within it, both in terms of densities and magnetic fields. These regions are nonetheless expected to be atypical, with strong radiation fields and magnetic turbulence, as well as fast outflows. 
In addition, the presence of a wind from the CMZ is now well established~\cite{BlandHawthorn03, DiTeodoro}, and advection may therefore dominate the transport of relativistic charged particles over a wide energy range in the GC.

The radiation field model used in this work is a self-consistent model of the broad-band continuum emission of our Galaxy and was derived from modelling maps of the all-sky emission in the infrared and submillimetre regime~\cite{radiationField}.
The magnetic field employed here follows the prescription of Jansson \& Farrar, including large-scale regular fields, striated fields and small-scale random fields~\cite{magnetic_field, Jansson_2012}. The regular field consists of a disk field and an extended halo field including a large, out-of-plane component and the orientation of the striated component is aligned with the regular field. The magnetic field topology is undoubtedly more complex in the inner \(\sim 100\) pc region \cite{Guendez}, though such complications are beyond the scope of the present work.

The diffusion coefficient in the region above the CMZ is poorly constrained at present, though there are strong physical arguments that indicate it is substantially smaller than the galactic average. TeV gamma-ray observations of the diffuse emission in the central region favour an hadronic origin, as evidenced by its close correlation with the gas maps \cite{HESS2018}. The resulting cosmic-ray distribution inferred from the gas maps is consistent with a $1/r$ radial profile, which constrains the diffusion coefficient to be substantially smaller than the galactic average, at least in the CMZ \cite{Scherer}. 

One can estimate a diffusion coefficient within the standard quasi-linear framework. The scattering rate of a magnetised particle, i.e. a particle undergoing helical motion about a mean field $B_0$, is (in cgs units) \cite[e.g.][]{Kulsrud},
\begin{equation*}
\nu =\left. \frac{\pi}{4} \left(\frac{k \mathcal{E}_k}{B_0^2/8\pi}\right) \frac{c}{r_{\rm g}}\right|_{k\mu r_{\rm g}=1},
\end{equation*}
where $r_{\rm g} = E/e B_0$ is the particle Larmor radius, $\mu$ its pitch angle with respect to the local mean magnetic field. $\mathcal{E}_k dk$ is the energy density in waves, with wavelengths $\lambda = 2\pi/k$, that are \emph{gyro-resonant} with particles of energy $E$ and pitch $\mu$. Assuming a turbulent power-law spectrum of scattering modes, $\mathcal{E}_k \propto k^{-\alpha}$, sharply cut-off above $\lambda > L$, an approximate expression for the average diffusion coefficient along $B_0$ is   
\begin{equation*}
D \lesssim 10^{27} \eta^{-1} \left\lbrace
\begin{array}{cl}
E_{\rm TeV}^{1/3} B_{10\mu{\rm G}}^{-1/3}L_{\rm pc}^{2/3}   
& \text{Kolmogorov} \\
E_{\rm TeV}^{1/2} B_{10\mu{\rm G}}^{-1/2}L_{\rm pc}^{1/2}     & \text{Kraichnen}
\end{array} \right.
\end{equation*}
Here, Kolmogorov and Kraichnen turbulent spectra correspond to $\alpha=5/3$ and $3/2$ respectively.
The quantity $\eta$ (assumed to be $\leq 1$) is the ratio of total integrated turbulent magnetic field energy density to that in the mean field, while $L_{\rm pc}$ is $L$ in units of parsecs, $E_{\rm Tev}$ the particle energy in TeV. At large galacto-centric radii $L_{\rm pc}\approx 100$ \cite[e.g.][]{BeckBeckBeck}. If we assume the turbulence in the inner regions is driven by stellar activity, $L$ should decrease as one moves closer to the CMZ. 
In the following, we set $\eta=1$ and take the outer scale of the turbulence $L$ to be $10\%$ of its distance from the Galactic Centre (i.e. approaching the average ISM value at 1 kpc). To simplify the calculations that follow, we take $D$ as a proxy for the radial diffusion coefficient and focus exclusively on Kolmogorov turbulence.  

\begin{figure}[htb!]
  \centering
  \includegraphics[width=\textwidth]{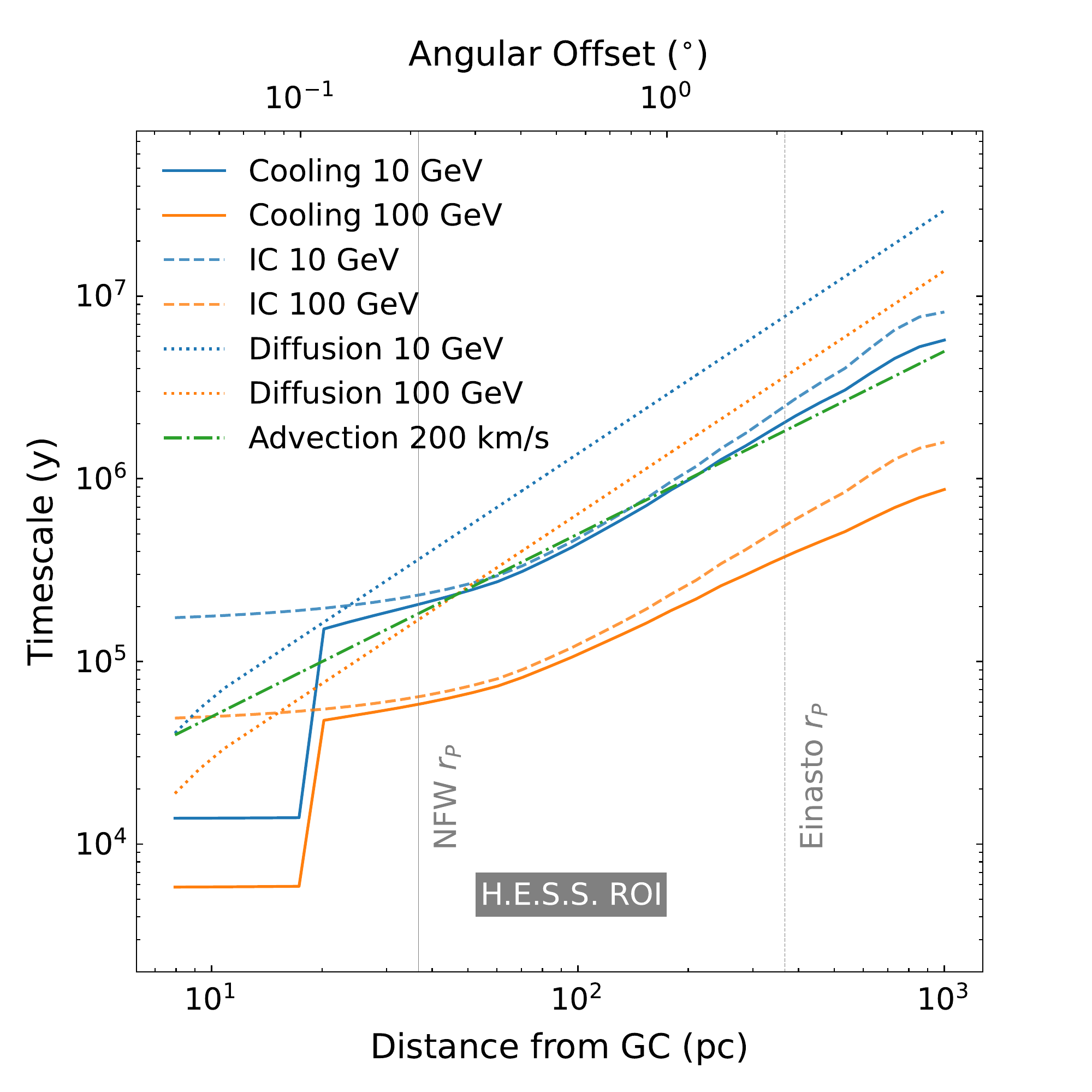}
  \caption{Timescales of electron cooling (solid lines), inverse Compton emission (dashed lines) and diffusion (dotted lines) as a function of the distance from the Galactic Centre. The timescales are shown for the two different electron energies 10\,GeV and 100\,GeV. The dash-dotted green line shows the timescale of advection with a velocity of 200\,km/s. The vertical lines show the radius at which the signal significance peaks for a DM distribution according to a NFW (solid line) and an Einasto (dashed line) profile. The grey area indicates a typical region of interest for  galactic indirect DM searches, here the one from~\cite{Abdallah_2016}.}
  \label{fig:timescales}
\end{figure}

With these assumptions about the conditions in the GC region, we use the open-source GAMERA package~\cite{GAMERA}, as employed e.g. in~\cite{Breuhaus_2022}, to calculate the timescales of the electron cooling. In Figure~\ref{fig:timescales} we compare them to the timescales of  electron propagation in the GC. The timescales are shown for electrons with two exemplary energies, namely 10 GeV (blue lines) and 100 GeV (orange lines). The total cooling timescales are shown by the solid lines, while the timescales of the inverse Compton emission are shown by the dashed lines. The dotted lines show the diffusion timescales for the different electron energies and the green dash-dotted line shows the advection timescales for a constant wind speed of 200\,km/s. 

Also shown are $r_{P}$, the radii with maximal signal significance\footnote{Where the signal significance corresponds to the ratio of the number of signal events (direct photon counts) and the square root of the number of background events for a constant background distribution within a sphere.}, for a dark matter distribution according to a Navarro, Frenk and White (NFW)~\cite{Navarro_1996, Navarro_1997} (grey solid line) and an Einasto~\cite{Einasto} (grey dashed line) profile. These radii correspond to the typical distances that are employed in galactic indirect DM searches as can be seen from the grey area that indicates the region of interest (ROI) of the galactic DM search of the H.E.S.S. Collaboration~\cite{Abdallah_2016}.

The shape of the WIMP distribution in the MW is however poorly constrained. While the cusp NFW and Einasto profiles have been widely employed in indirect DM searches, there is experimental evidence that prefers shallower cusps or even cored density profiles~\cite{Portail16}. The latter profiles have larger values for $r_{P}$, which means that the expected signal strength is weaker in the GC than for the cusp profiles and that the shallow cusp or cored morphology is disadvantageous for WIMP searches at the GC.

From Figure~\ref{fig:timescales} it is evident that for the GC distances that are explored by gamma ray telescopes the IC emission dominates the cooling. At larger distances synchrotron effects become more important while at smaller distances both synchroton emission and Bremsstrahlung contribute significantly to the electron cooling (not shown in Figure~\ref{fig:timescales}). At about 20\,pc the synchroton emission and Bremsstrahlung lead to a sudden step in the cooling timescales and at even smaller distances the cooling times are only tentative as the magnetic field model breaks down inside the CMZ. 

All in all, we can see from Figure~\ref{fig:timescales} that the cooling timescales are shorter than the propagation timescales for energetic electrons ($>$ 10\,GeV) meaning that the electrons are more likely to cool locally/in-situ than to leave the GC region.

\section{Inverse Compton Emission}
\label{sec:IC}
Electrons that are produced via the annihilation of heavy WIMPs in the GC mostly cool locally/in-situ as shown in Figure~\ref{fig:timescales}. This energy loss is likely dominated by inverse Compton scattering especially in the region that is investigated by DM searches with IACTs.
During the IC process electrons scatter off photons from the CMB, starlight, UV and IR light and transfer their energy such that energetic photons are produced. Consequently, the electrons from WIMP annihilation in the GC give rise to an additional component in the photon spectrum apart from the photons originating directly from the annihilation process, from the same spatial region as the direct signal. 

For our estimations, we use the spectra of electrons and positrons from WIMP annihilation described in the widely used 'cookbook'~\cite{COOKBOOK} and available in numerical form~\cite{pppc4dmid_website, COOKBOOK_ewCorr}. For the calculation of the resulting photon spectra, we employ the GAMERA package. We assume a constant injected power over a given timescale, resulting in an equilibrium situation above a certain energy. The WIMP density profile governs at which position the IC emitting particles are produced but 
the results presented here are largely independent of the density profile, at least in the energy range where in-situ/local cooling can safely be assumed.

\begin{figure}[htb!]
 \begin{subfigure}[b]{0.5\textwidth}
         \centering
         \includegraphics[width=\textwidth]{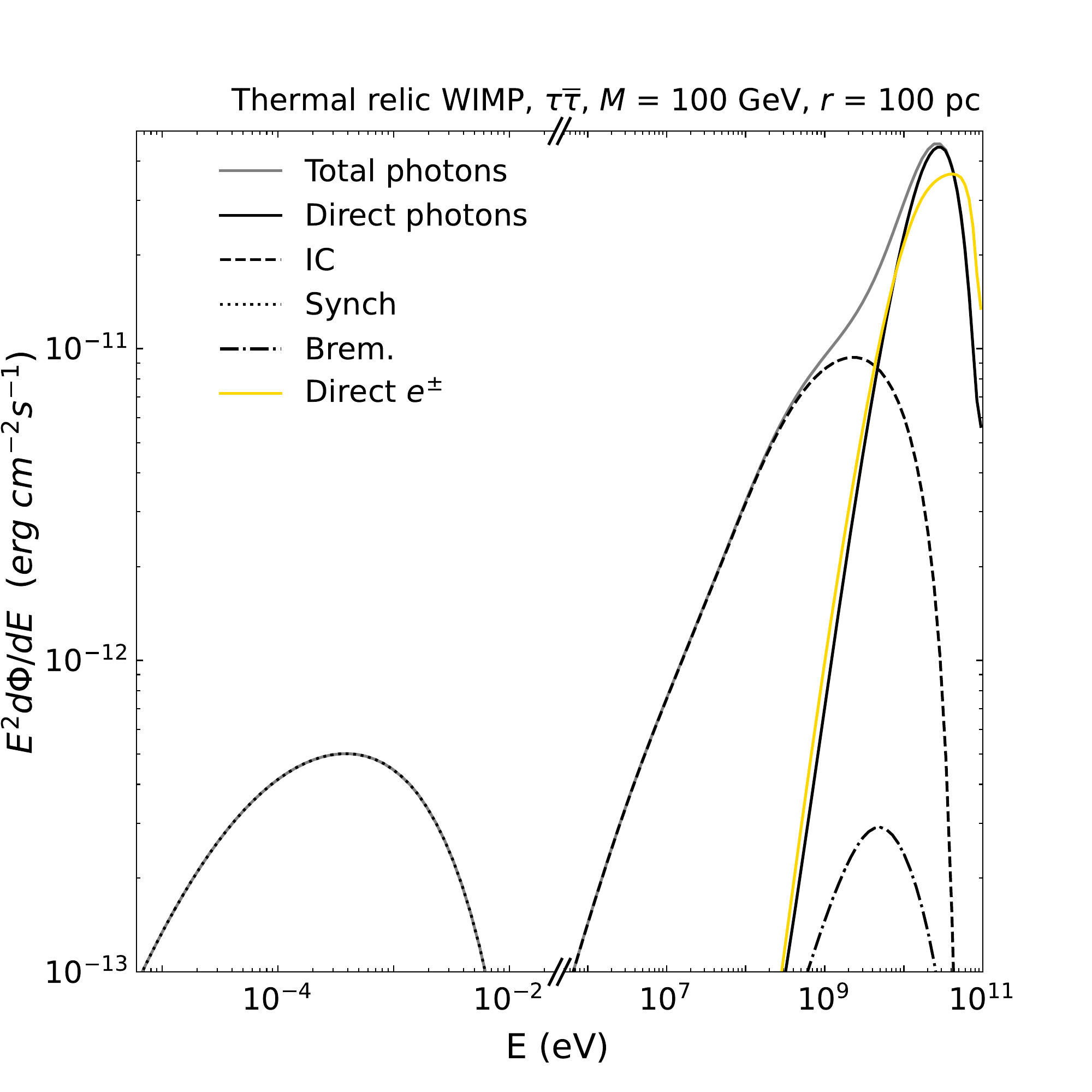}
         \caption{Total photon spectrum and its components}
         \label{fig:total}
     \end{subfigure}
     \begin{subfigure}[b]{0.5\textwidth}
         \centering
         \includegraphics[width=\textwidth]{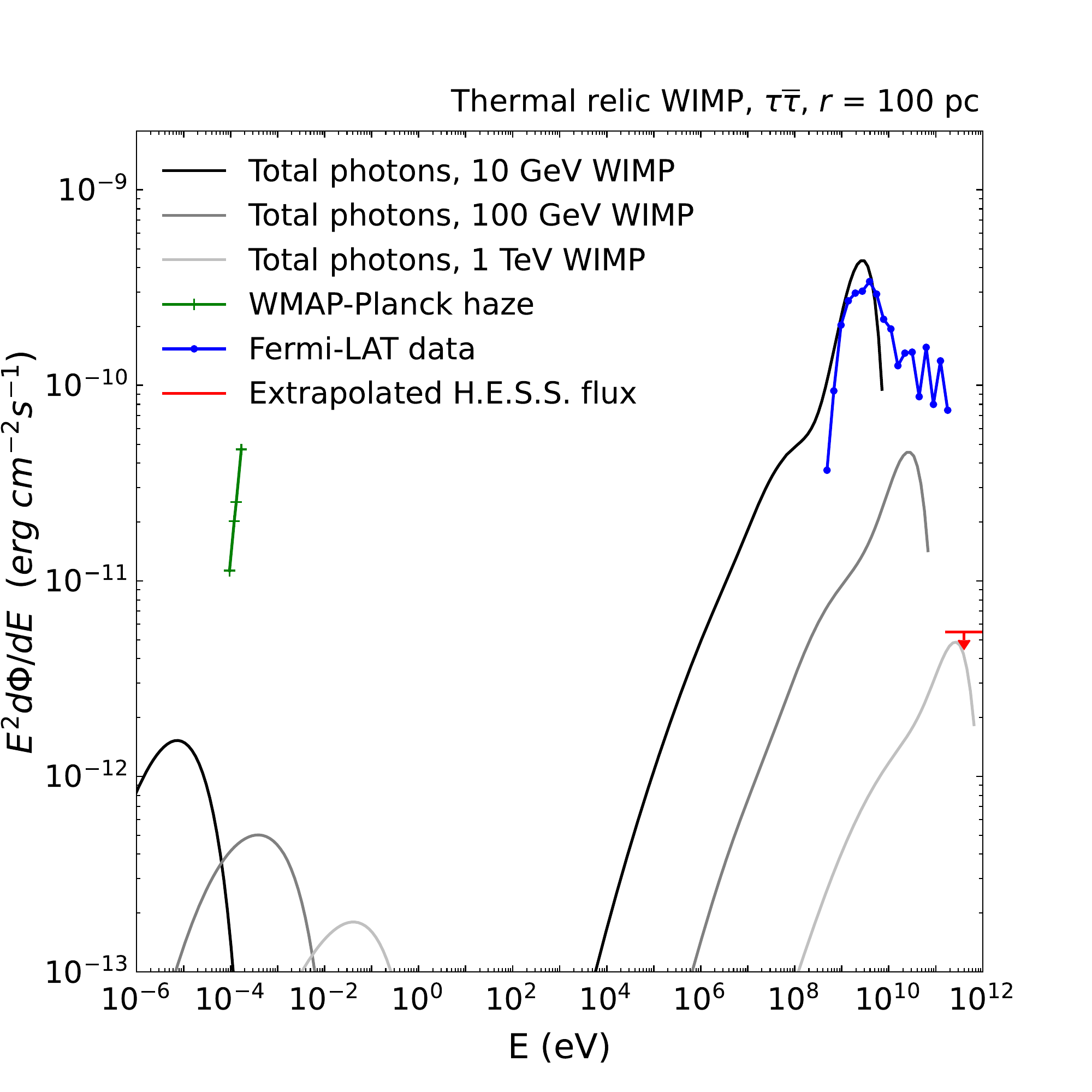}
         \caption{Comparison of expected spectra with data}
         \label{fig:dataComp}
     \end{subfigure}  
    \caption{Total photon flux from WIMP annihilation to $\tau$ leptons in the GC. For the normalisation, the J-factor is set to $1.53 \times 10^{22}$\,GeV$^{2}$cm$^{-5}$ and the thermal cross section of $\langle \sigma v \rangle = 3 \times 10^{-26}$\,cm$^{3}$/s is used. Left panel: The gray solid line shows the total photon spectrum that is composed of the direct photon component (solid black line) and the indirect photon components: Inverse Compton emission, synchrotron radiation and Bremsstrahlung (dashed, dotted and dash-dotted line respectively). The yellow line shows the spectrum of the electrons and positrons that are directly emitted from the WIMP annihilation and that give rise to the indirect photon contributions. The WIMP mass is set to 100\,GeV and the model was evaluated at a distance of 100\,pc from the GC.
    Right panel: Comparison of the total photon spectra for 3 different WIMP masses with experimental data. The WMAP haze measured with Planck~\cite{planck_2013} (green), the Fermi GC excess~\cite{Fermi_2017} (blue) and the extrapolated flux measured with H.E.S.S.~\cite{Abdallah_2016} (red).}
  \label{fig:abs}
\end{figure}

Figure~\ref{fig:abs} shows the differential electron and photon flux times squared energy (i.e. total power output per logarithmic energy interval) expected for WIMP annihilation to $\tau$ leptons. The direct photon spectrum (black solid line in the left panel) is taken from~\cite{COOKBOOK, pppc4dmid_website,  COOKBOOK_ewCorr} and corresponds to a model-independent baseline photon contribution, i.e. three-body final states of WIMP annihilation are omitted -- even though they might be viable in some models.
The component denoted ``direct $e^{\pm}$'' (yellow line) corresponds to twice the tabulated positron flux from~\cite{COOKBOOK, pppc4dmid_website,  COOKBOOK_ewCorr} to account for electron as well as positron production. This direct electron-positron spectrum is used as input to our model and evaluated in this case at a distance of 100\,pc from the GC as this is the typical region of interest for gamma-ray WIMP searches (as illustrated in Figure~\ref{fig:timescales}). The model is evolved for $10^{6}$\,years and the escape of low energetic electrons due to advection is implemented in GAMERA in order to obtain the indirect components of the photon spectrum (broken lines). The indirect photon components add a low energy tail to the direct component. The former is dominated by the IC emission for $>$GeV-mass WIMPs which is why we concentrate on this contribution in this work. Bremsstrahlung is not expected to contribute significantly in the low density region outside of the Galactic disc / CMZ, synchrotron radiation contributes at much lower energies. 

The total photon spectrum is given by the sum of all broken lines in the left panel of the figure and illustrated as a gray solid line. The gray curves in the left and right panel are equivalent and show the expected photon spectrum for a thermal relic WIMP with $\langle \sigma v \rangle = 3 \times 10^{-26}$\,cm$^{3}$/s and a mass of 100\,GeV. The branching ratio to $\tau$ leptons is set to 1. For the normalisation, a $J$-factor of $1.53 \times 10^{22}$\,GeV$^{2}$cm$^{-5}$ is used and corresponds to the value computed in~\cite{Fermi_2017} for a NFW profile.  

In the right panel of Figure~\ref{fig:abs}, the expected photon spectra are compared to experimental data to illustrate the respective constraining power. The green curve corresponds to the WMAP haze measured with Planck~\cite{planck_2013} in a field of view of 10\degree\ and constrains the synchrotron contribution of the WIMP annihilation spectra. In blue the data points of the so-called Fermi GC excess~\cite{Fermi_2017} are shown. These data constrain thermal relic WIMPs up to tens of GeV in mass and is particularly sensitive to the indirect component of the photon spectrum. The red line indicates the flux that should have been measured with H.E.S.S. if its ROI corresponded to the one of Fermi. To extrapolate this value, the expected direct photon spectrum of~\cite{COOKBOOK} for a 1\,TeV WIMP is used together with the excluded $\langle \sigma v \rangle$ value from~\cite{Abdallah_2016}. To account for the difference in size between the Fermi and the H.E.S.S. ROI, the computed H.E.S.S. flux is scaled by the ratio of the J-factors for NFW profiles from~\cite{Fermi_2017} and~\cite{Abdallah_2016}. The H.E.S.S. data excludes thermal relic WIMPs with TeV mass for this annihilation channel and assumed profile. This figure illustrates that Galactic WIMPs can be probed in multiple wavelength ranges if all emission components are taken into account and that the direct photon component covers only a comparatively narrow energy band. 

\begin{figure}[ht]
 \begin{subfigure}[b]{0.5\textwidth}
         \centering
         \includegraphics[width=\textwidth]{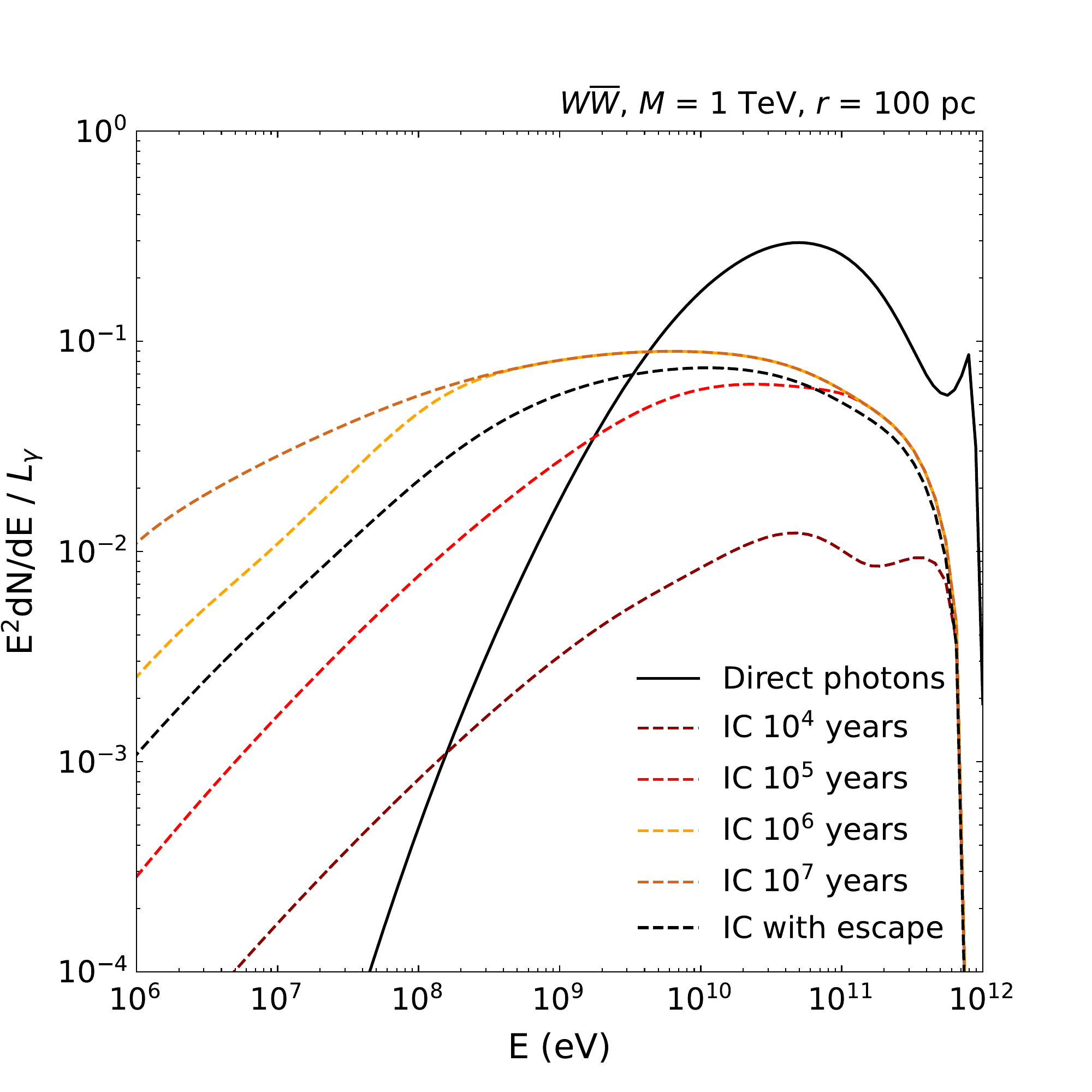}
         \caption{Time dependence of IC emission}
         \label{fig:times}
     \end{subfigure}
    % \hfill
     \begin{subfigure}[b]{0.5\textwidth}
         \centering
         \includegraphics[width=\textwidth]{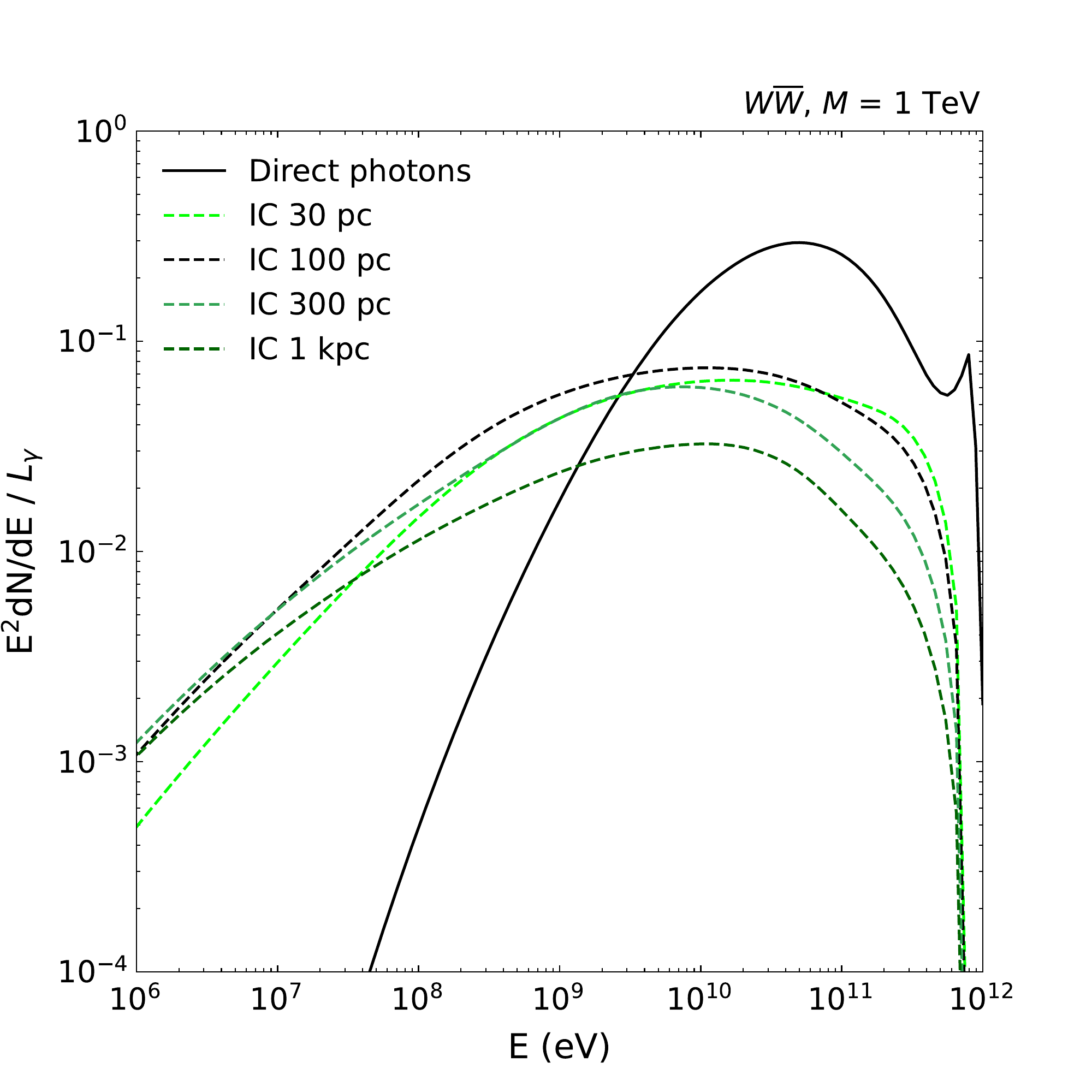}
         \caption{Distance dependence of IC emission}
         \label{fig:distances}
     \end{subfigure}
\caption{Spectrum of direct and inverse Compton photons from WIMP annihilation to $W$ bosons in the GC, normalised to the total luminosity of the direct photons. The solid line shows the spectrum for direct photons while the dashed lines show the spectra of IC photons. The latter are shown for different timescales of ongoing DM annihilation in the left panel (at 100 pc distance from the GC) and for different distances from the GC in the right panel (for a fixed timescale of $10^{6}$ years). The WIMP mass is set to 1\,TeV in this example.}
\centering
\label{fig:timeNdistance}
\end{figure}

Having established the nature and importance of the IC component of the photon spectrum, we are investigating its robustness in the remaining part of this section. In Figure~\ref{fig:timeNdistance} the evolution of the IC component of the photon spectrum from a WIMP signal is shown as a function of time (left panel) and as a function of distance (right panel). The spectra are normalised to the total luminosity of the direct photons, $L_{\gamma}$. 

From Figure~\ref{fig:timescales} it is clear that the relevant timescales for residence of injected particles in the Galactic Centre region ranges from $10^{5}$ -- $10^{7}$ years. Figure~\ref{fig:times} shows the impact of this confinement timescale on the resulting spectral energy distribution of IC emission. The coloured components show  the IC emission of the electrons after evolving our model for a given time. The black dashed line illustrates the nominal implementation of our model and corresponds to an evolution of the system for $10^{6}$\,years including the escape of low energetic electrons due to advection. By accounting for the escape of electrons, the timescale of the evolution of the model becomes irrelevant as long as it is larger than $10^{6}$\,years. 
On very short timescales the emission is dominated by the highest energy electrons, and the illustrated case of a 1 TeV WIMP ($W\overline{W}$ channel), significantly suppressed by the Klein-Nishina (KN) effect. 
On the timescales relevant for electrons in the GC ( $> 10^{5}$ years) the IC emission constitutes a significant component of the total photon spectrum. 

The flux of the IC component of the photon spectrum depends on the local ratio of magnetic to radiation energy density. The spectral shape of the IC component depends on the local radiation spectral energy distribution (SED). Figure~\ref{fig:distances} shows the dependence of the resulting IC SED on the distance from the GC within the framework of our model, for the same injected spectrum as in Figure~\ref{fig:times}. Overall the impact of changing conditions is relatively modest, provided the distance of the emission region from the GC is $<$1 kpc -- as expected for the bulk of the annihilation signal for almost any reasonable density profile. 

\begin{figure}[ht]
 \begin{subfigure}[b]{0.5\textwidth}
         \centering
         \includegraphics[width=\textwidth]{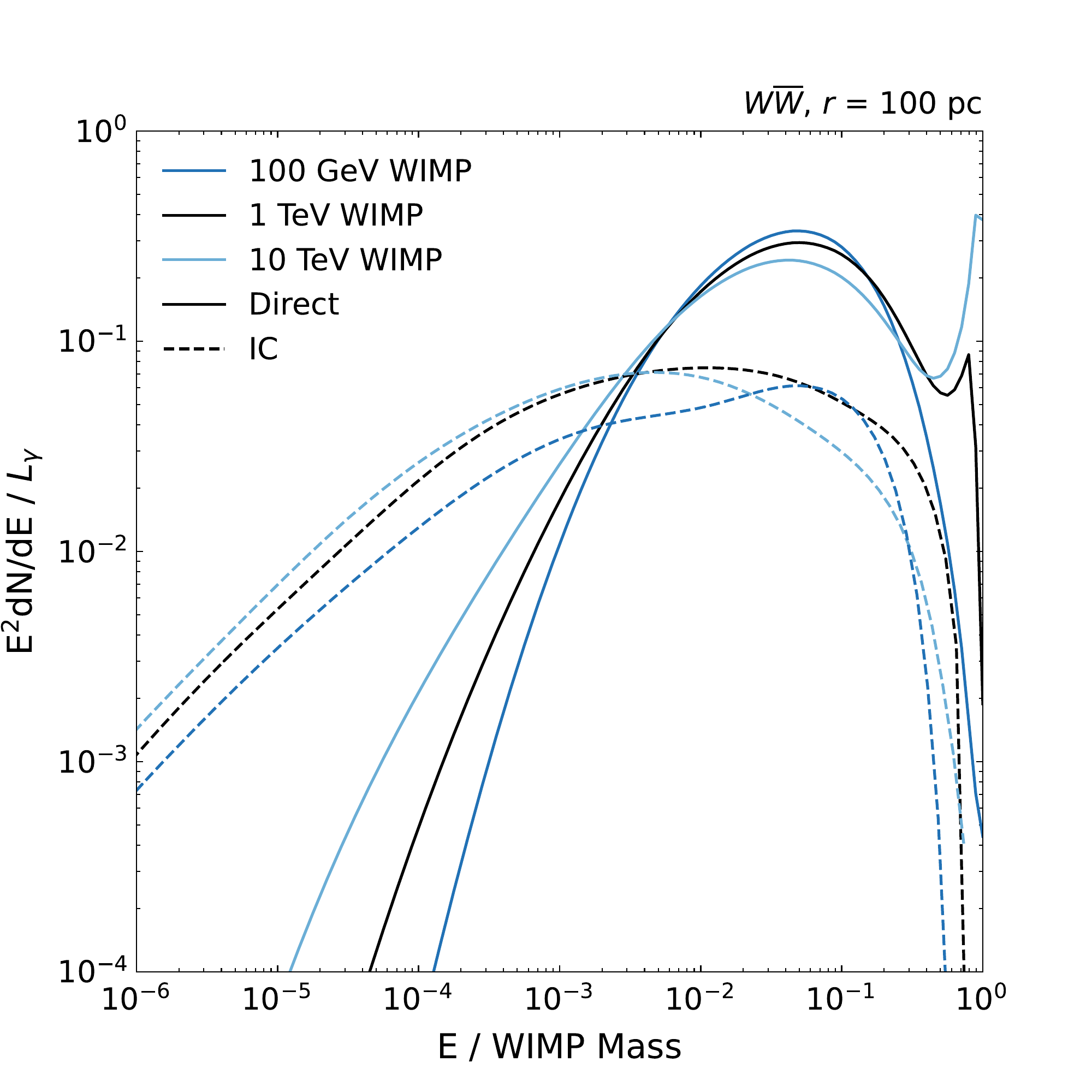}
         \caption{WIMP mass dependence of IC emission}
         \label{fig:masses}
     \end{subfigure}
     %\hfill
     \begin{subfigure}[b]{0.5\textwidth}
         \centering
         \includegraphics[width=\textwidth]{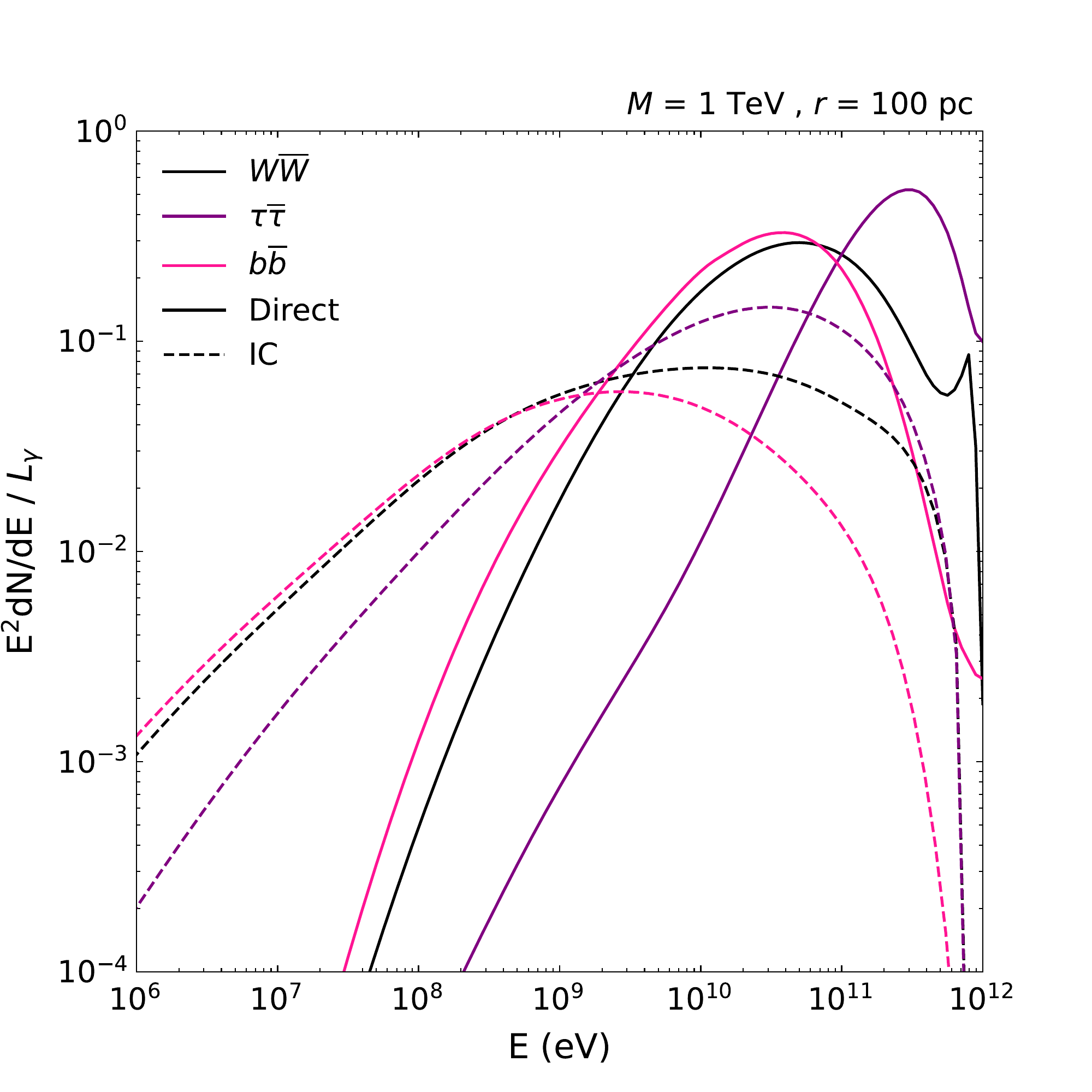}
         \caption{Decay channel dependence of IC emission}
         \label{fig:channels}
     \end{subfigure}
\caption{Spectrum of direct and inverse Compton photons from WIMP annihilation in the GC normalised to the total luminosity of the direct photons. The solid lines show the spectrum for direct photons while the dashed lines show the spectra of IC photons. In the left panel the spectra are shown for different WIMP masses in the $W$ annihilation channel, while the right panel shows the spectra for different annihilation channels and a WIMP mass of 1\,TeV.}
\centering
\label{fig:massNchannel}
\end{figure}

In Figure~\ref{fig:massNchannel} the direct and IC components of the photon spectra are compared for different WIMP masses and decay channels. Again, the spectra are normalised to the total luminosity of the direct photons. The model is evaluated at 100\,pc from the GC after an evolution time of $10^{6}$\,years and includes electron escape with the nominal outflow velocity as illustrated in Figure~\ref{fig:timescales}.

Changing the mass of the annihilating WIMP modifies both the spectrum of injected electrons and the impact of KN suppression on the resulting IC emission. With electroweak corrections, the $W\overline{W}$ channel from~\cite{COOKBOOK} produces harder photon and electron spectra at higher masses, at the same time the emission from injected high-energy electrons is increasingly suppressed by the KN effect. As shown in Figure~\ref{fig:masses}, the result of these two competing effects is a relatively stable IC spectrum over a wide mass range (0.1--10\,TeV).

Considering different production channels for electrons and photons, however, has a dramatic effect on the resulting IC emission (see Figure~\ref{fig:channels}). The very hard photon and electron production associated to the $\tau\overline{\tau}$ channel results in a high peak energy and reduction in low energy IC emission in this case. The soft spectra produced in the $b\overline{b}$ channel result in correspondingly soft IC emission at high energies. In all cases however,  the IC emission represents a significant low energy shoulder to the total gamma-ray emission of the GC region due to WIMP annihilation. 

\begin{figure}[htb!]
  \centering
  \includegraphics[width=.65\textwidth]{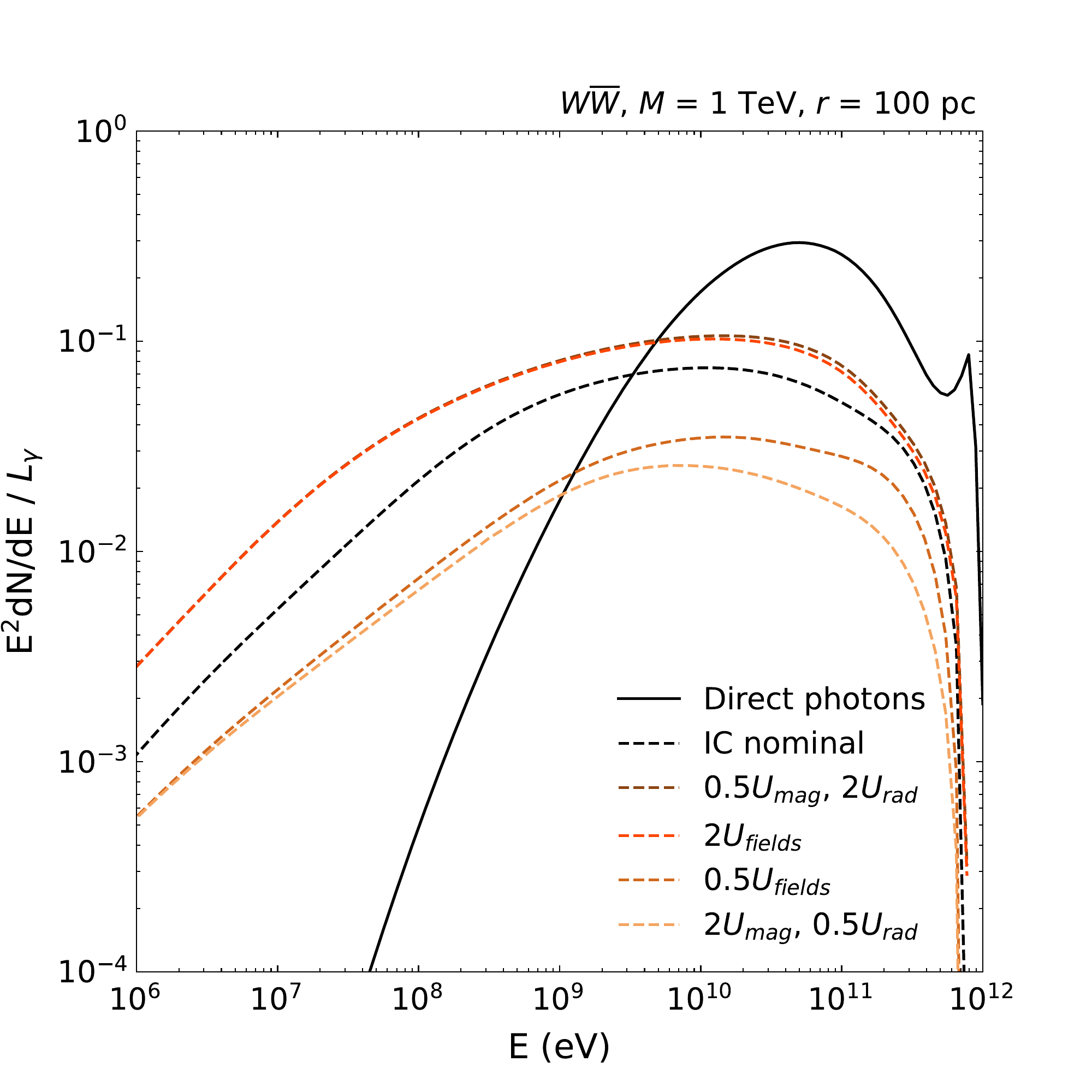}
  \caption{Photon spectrum from WIMP annihilation to $W$ bosons in the GC normalised to the total luminosity of the direct photon component (solid line). The dashed lines show the IC component of the spectrum for a variation of the magnetic field strength and the radiation field of the model. The WIMP mass is set to 1\,TeV and the model was evaluated at a distance of 100\,pc from the GC and evolved for $10^{6}$\,years.}
  \label{fig:conditions}
\end{figure}

The stability of the IC component under a change in the model is illustrated in Figure~\ref{fig:conditions}. The different colours correspond to different energy densities of the magnetic field and the radiation field model. They are varied by a factor of 2 from the nominal value to illustrate the uncertainty associated with these components.

\section{Discussion}
Based on the results presented here, the presence of significant additional low energy photons associated to IC emission in the GC from annihilating Dark Matter appears to be a robust prediction for $\sim$TeV mass WIMPs. As such it should be incorporated in future experimental searches. A particular concern is that searches increasingly make use of a template spectrum and a `3D' likelihood fitting approach, and in neglecting the IC component may reach misleading conclusions. 

\begin{figure}[ht!]
 \begin{subfigure}[b]{0.5\textwidth}
         \centering
         \includegraphics[width=\textwidth]{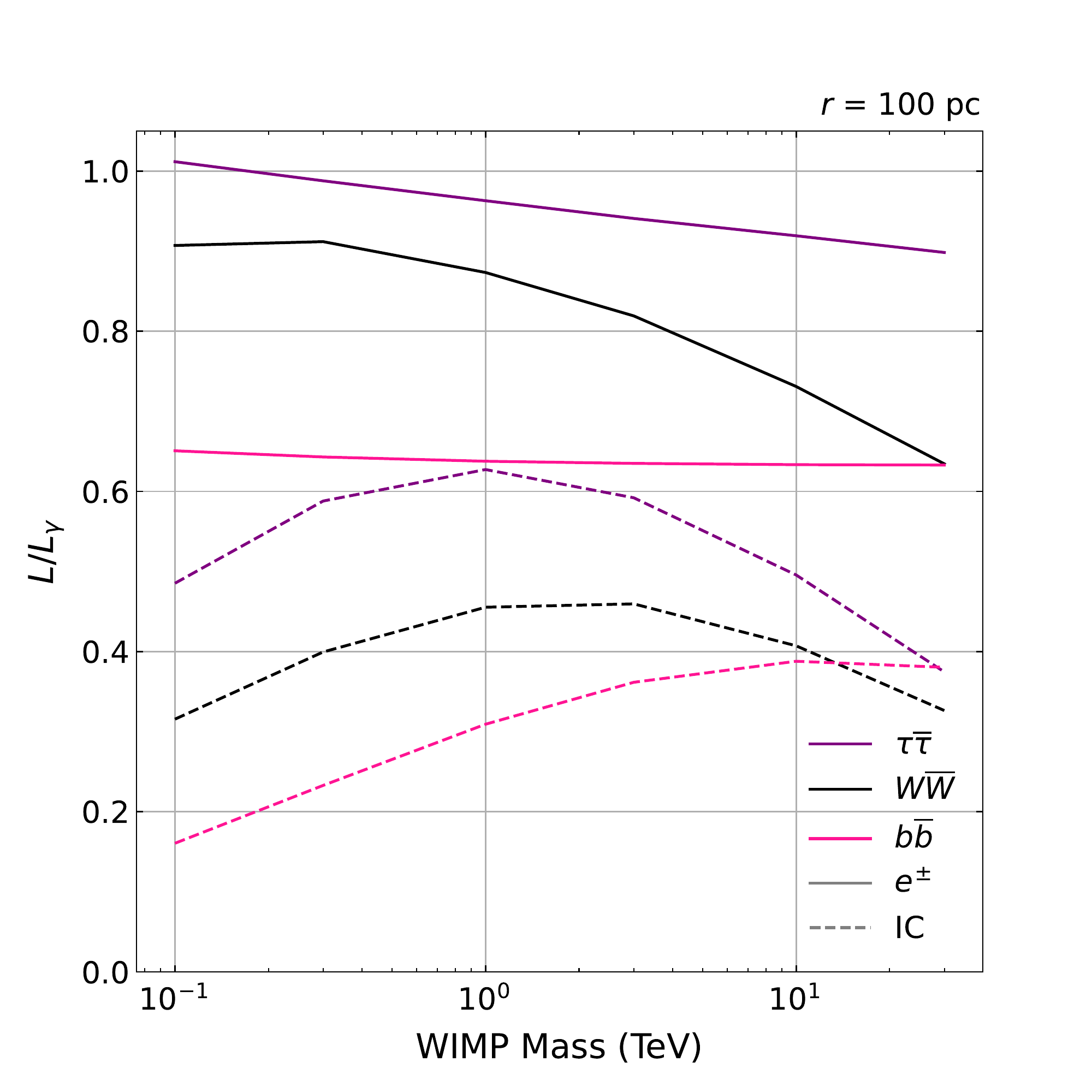}
         \caption{Relative production efficiency}
         \label{fig:efficiency}
     \end{subfigure}
     %\hfill
     \begin{subfigure}[b]{0.5\textwidth}
         \centering
         \includegraphics[width=\textwidth]{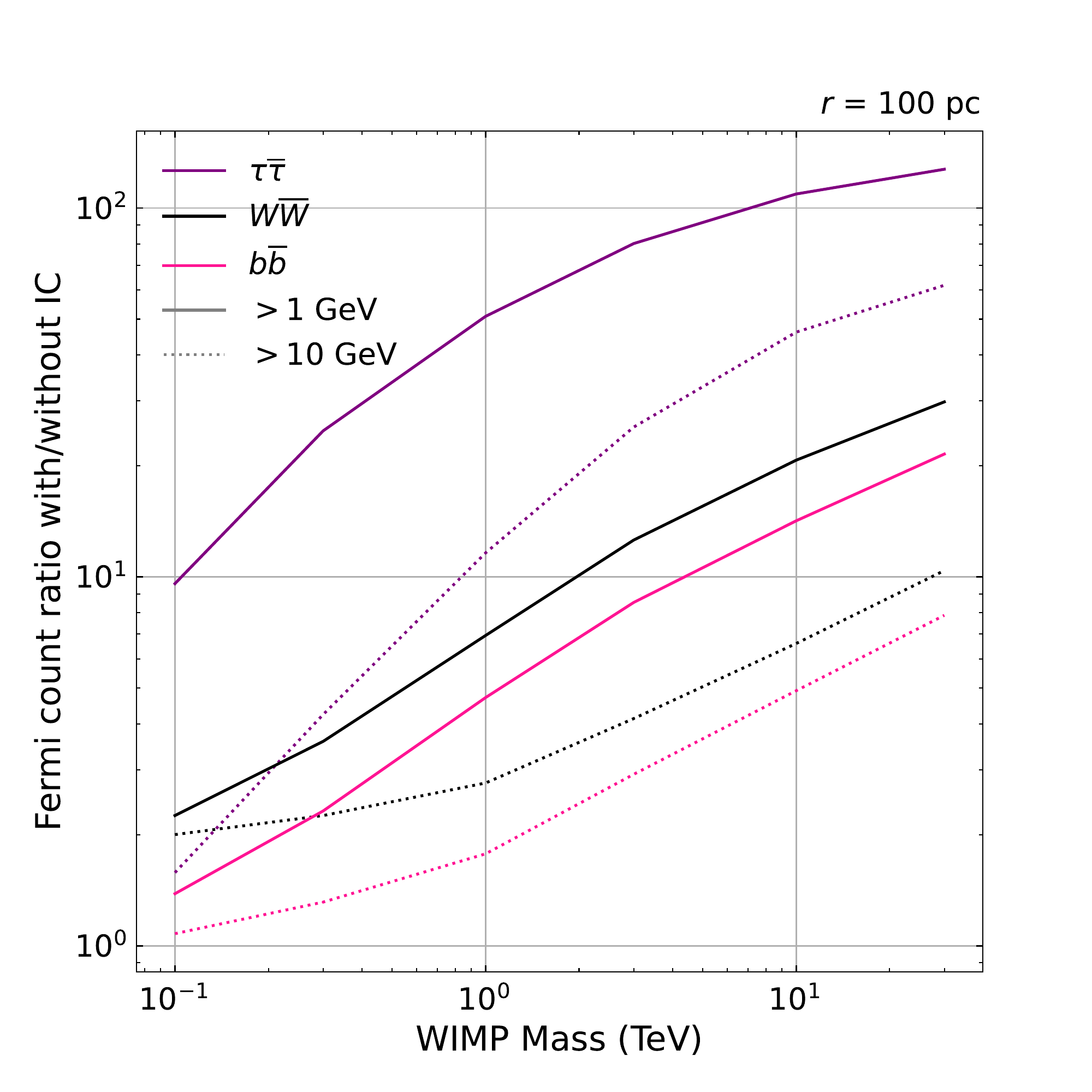}
         \caption{Signal count ratio with/without IC photons}% for the Fermi-LAT}
         \label{fig:fermi}
     \end{subfigure}
\caption{Left: Overall power input to electrons from WIMP annihilation relative to the overall power input to direct photons (solid lines). The dashed lines show the same quantity for the resulting IC emission. 
Right: Ratio of the photon counts of a DM signal with and without taking the IC component into consideration as a function of the WIMP mass. The solid (dotted) lines correspond to an energy threshold of the experiment of 1\,GeV (10\,GeV).
Three different WIMP annihilation channels $W\overline{W}$, $\tau\overline{\tau}$ and $b\overline{b}$ are shown in black, purple and violet, respectively.
}
\centering
\end{figure}

As mentioned above, there is a competition between hardening production spectra and increasing KN suppression as WIMP masses get heavier. 
Figure~\ref{fig:efficiency} shows the overall power input in to electrons (solid lines) and subsequently in to IC emission (dashed lines), relative to the power of the direct photon component. The three different decay channels $W\overline{W}$, $\tau\overline{\tau}$ and $b\overline{b}$ are shown. The figure shows that not all the power of the electrons are put into the IC emission and that for extremely heavy ($\sim$30 TeV) WIMPs, the IC power starts to be quite strongly suppressed. Still, the effect of IC emission is certainly relevant over a very wide mass range in all production channels.

The impact that the IC emission can have on detectability is strongly dependent on the energy-dependent sensitivity of the gamma-ray detector considered. For Fermi-LAT the energy-flux sensitivity reaches a maximum in the GeV region, which is why the IC component does have a considerable impact on the expected signal counts for heavy WIMPs. Whilst a complete DM search with Fermi-LAT, including the evaluation of background effects, is beyond the scope of this work, we estimate the increase in the signal size seen in Fermi-LAT with and without the IC component, in the case of heavy WIMPs. To this end, we parameterise the acceptance of the Fermi-LAT Pass 8 data Release 3 ({\it UltraCleanVeto}) Version 3~\cite{FermiWebpages} and estimate the ratio of expected photon counts with and without taking the IC component into account. This ratio is shown as a function of the WIMP mass in Figure~\ref{fig:fermi}. The $W\overline{W}$, $\tau\overline{\tau}$ and $b\overline{b}$ annihilation channels are shown in black, purple and violet, respectively. The solid lines show the expected count ratio for an energy threshold of 1\,GeV and the dotted lines correspond to an energy threshold of 10\,GeV. The figure shows that including the IC component of the WIMP signal, increases the expected signal yield for Fermi-LAT significantly -- in the case of heavy WIMPs decaying to $\tau$ leptons by more than a factor 100. Therefore, we conclude firstly that the current Fermi-LAT limits that disregard the IC component are too pessimistic and would improve when adding the IC component to the signal model. Secondly, due to the existence of the IC component, the Fermi-LAT data could confirm or challenge a potential claim of heavy WIMP observation from future ground-based gamma-ray telescopes, making the data an important asset for DM searches for many years to come.

The impact of the IC component of the WIMP signal on ground based gamma-ray observatories is expected to be smaller than the effect it has for Fermi-LAT, except in the case of extremely high mass WIMPs.

This work provides a first estimate of the IC component of the gamma-ray signal from WIMP annihilation in the GC. More sophisticated models of the GC environment should be considered in the future, as well as the intermediate regime ($\sim$1-10 GeV electrons) where particles are  transported significant distances before cooling, with an impact on the measured emission profile.

Finally we note that any claim of a gamma-ray WIMP annihilation signal must be tested against radio-mm limits on the accompanying synchrotron emission. Although, as Figure~\ref{fig:dataComp} shows, synchrotron constraints for a thermal relic are expected to be relatively weak unless the GC magnetic field strength is much higher than the model adopted here.

\section{Conclusions}
The gamma ray spectrum of WIMP annihilation in the centre of our galaxy consists of two components originating from direct and indirect photons respectively. We have shown that the latter component is dominated by IC emission for heavy WIMPs as the electrons produced by the annihilation radiate locally, at least at sufficiently high energies. Whilst uncertainties related to transport and cooling are significant, they do not exceed the fundamental uncertainty due to the currently poorly constrained DM halo profile of the Milky Way. We evaluated the robustness of the IC component for different timescales, distances from the GC, WIMP masses and production channels. All in all, we found that the IC component is significant and should not be neglected in WIMP searches.

In addition, we evaluated the effect the IC component has on the detectability of heavy ($\sim$TeV) WIMPs with Fermi-LAT. Including the IC component significantly strengthens the WIMP signal and revised limits from the Fermi collaboration incorporating IC emission are desirable. 

In order to solve the long standing mystery of the particle nature of DM all aspects of the signal should be taken into account. We therefore urge that also the indirect component of the gamma-ray spectrum should be considered in the next generation of Galactic WIMP searches with gamma-ray observatories. This is especially true in case a WIMP signal is observed, as the shape of the spectrum will give crucial hints on the nature of the particle.

\section*{Acknowledgements}
J.D., is funded by the Research Council of Norway, project number
301718 and grateful for the hospitality of the Non-Thermal Astrophysics division at MPIK.


\bibliographystyle{elsarticle-num} 
\bibliography{icGCphotons}
%\begin{thebibliography}{00}
%\bibitem{}
%\end{thebibliography}

\end{document}